\documentclass[namedreferences]{SolarPhysics}
\usepackage[optionalrh]{spr-sola-addons} % For Solar Physics 
\usepackage{graphicx,rotating}        % For eps figures, newer & more powerfull
\usepackage{amssymb}        % useful mathematical symbols
\usepackage{color}           % For color text: \color command
\usepackage{url}             % For breaking URLs easily trough lines
\begin{document}

\begin{article}

\begin{opening}

\title{Newly identified properties of surface acoustic power}

\author{H.~\surname{Schunker}$^{1}$\sep
        D.C.~\surname{Braun}$^{2}$ 
       }
\runningauthor{H. Schunker \textit{et al.}}
\runningtitle{Properties of surface acoustic power}

   \institute{$^{1}$ Max Planck Institute for Solar System Research, Max-Planck 
Strasse 2, Katlenburg-Lindau 37191, Germany
                     email: \url{schunker@mps.mpg.de} \\ 
                    $^{2}$ NorthWest Research Associates, Inc., Colorado 
Research Associates Division, 3380 Mitchell Lane, Boulder, CO 80301, U.S.A.\\
             }

\begin{abstract}
The cause of enhanced  acoustic power surrounding active regions, the acoustic 
halo, is not as yet understood. We explore the properties of the enhanced 
acoustic power observed near disk center from 21 to 27 
January 2002, including AR 9787. We find that (i) there exists a strong correlation of the enhanced high 
frequency power with magnetic-field inclination, with greater power in more horizontal fields, (ii) the frequency of  the maximum enhancement increases along with magnetic field strength, and (iii) the 
oscillations contributing to the halos show modal ridges which are shifted to 
higher wavenumber at constant frequency in comparison to the ridges of modes in the 
quiet-Sun.
\end{abstract}
\end{opening}
%-------------------------------------------------

\section{Introduction}
     \label{introduction}
      
     The application of local helioseimology to probe the nature of solar active regions
is critically important in understanding their origin and evolution. The influence
of magnetic fields on incident acoustic waves is complex, but is coming under
increased scrutiny due to advances in numerical simulations of wave propagation in
magnetised plasmas (e.g. \opencite{KC06}; \opencite{SET06}; \opencite{PK07}; \opencite{CGD08}; \opencite{RSK09}). A successful model should reproduce all of the
observed features, including changes in the surface amplitudes of waves in the vicinity
of active regions.

The acoustic halo is an observed enhancement of high frequency
acoustic power surrounding regions of strong magnetic field.
This enhancement was discovered independently by \inlinecite{BLFJ92}, 
\inlinecite{BBLT92} and \inlinecite{TL93}. Observations
made using Ca II K-line images showed these enhancements extending tens of
Mm beyond active regions \cite{BLFJ92, TL93}. 
\inlinecite{BBLT92} showed patchier and more confined halos using
Doppler velocity
measurements taken in the Fe I $\lambda$5576 line, and with higher spatial
resolution. They
 also showed a peak in the power at 5.5~mHz.
\inlinecite{Hindman:1998p398} and later
\inlinecite{JH02} demonstrated that halos seen in SOHO/MDI \cite{MDI95} Dopplergrams
tend to be prominent in intermediate magnetic field strengths of 50-250 G and 
are absent in continuum intensity observations.
Acoustic halos  have been shown to be more prominent in Ca II K-line power maps 
and less visible in the Doppler Ni I line shifts \cite{Ladenkov:2002p369}.  
 \inlinecite{Finsterle2004} found that the acoustic halo increases in size with 
increasing height into the canopy over active regions. 
On the other hand, \inlinecite{M05} found that high-frequency oscillations in TRACE  
UV filtergrams show deficits in magnetic fields, rather than enchancements.
There is no established explanation for the enhanced power, although several
ideas have been suggested (see Section~\ref{discussion}).

 %what we present
In this paper we present an analysis revealing several new properties of acoustic halos
and of surface acoustic power in general. We use data obtained
by SOHO/MDI over 7 days containing
several large active regions. 
In Section~\ref{one} we demonstrate (i) 
a strong association of the enhanced high frequency power with nearly 
horizontal field, and (ii) 
an increase in the temporal frequency of the maximum enhancement with field 
strength. In Section~\ref{two} we show that (iii) the 
oscillations contributing to the halos show modal ridges which are shifted to 
higher wavenumber at constant frequency than the ridges due to waves in the 
quiet-Sun. 
In the Discussion we list possible theories put forward, to date, that attempt to 
explain the existence of the acoustic halos, noting that none of them have been shown to explain all 
of the observed properties successfully.

 \section{Correlations between the acoustic power enhancement and the magnetic field} 
\label{one}%%%%%%%%%%%%%%%%%%%%%%%%%%%%%%%%%%%%%%%%

We use 7 days of observations from 21 - 27 
January 2002 covering a significant number of active regions including
  AR 9787 which has already been subjected to a comprehensive helioseismic analysis 
\cite{Gizon:2009p660}.
We track the magnetograms and Dopplergrams, remapped into Postel's projections, 
in 24 hour blocks centered at the 
disk center for the middle of that day. The scale of the map is 
$0.002 R_\odot \approx 1.4$~Mm per pixel over 700 pixels, so the extent of the 
maps are about 975~Mm in each direction. The acoustic power maps are 
normalised to the quiet-Sun power by removing a spatial 2D 
4th degree polynomial fit to the quiet-Sun power for each day and frequency band. 
This fit removes several effects including center-to-limb variations
in the observed mode amplitudes and spatial fluctuations in the instrumental MTF. The latter
includes variations of focus across the CCD. 

The quiet-Sun is defined to be those pixels where the total magnetic field strength, $|B|=\sqrt{B_x^2 + B_y^2 + B_z^2}$,
is less than 30~G. Standard (potential field) methods were applied to reconstruct the vector magnetic field, $(B_x, B_y, B_z)$, at the Sun's surface from the line-of-sight components given by the SOHO/MDI observations (see \inlinecite{S82} for the more general problem of extrapolating the field into the overlying corona). The $x$-direction is defined as positive in the direction of solar East perpendicular to the meridian, the $y$-direction is defined as positive towards solar South perpendicular to the equator and the radial magnetic field is represented by $B_z$.   Keeping in mind that potential field 
extrapolations are not the best representation of the actual magnetic field 
\cite{Wiegelmann:2005p693}, we avoid using the strong fields (for example, in sunspots) where 
such extrapolations are known to be particularly unreliable.

\begin{figure}
\includegraphics[width=1.0\textwidth]{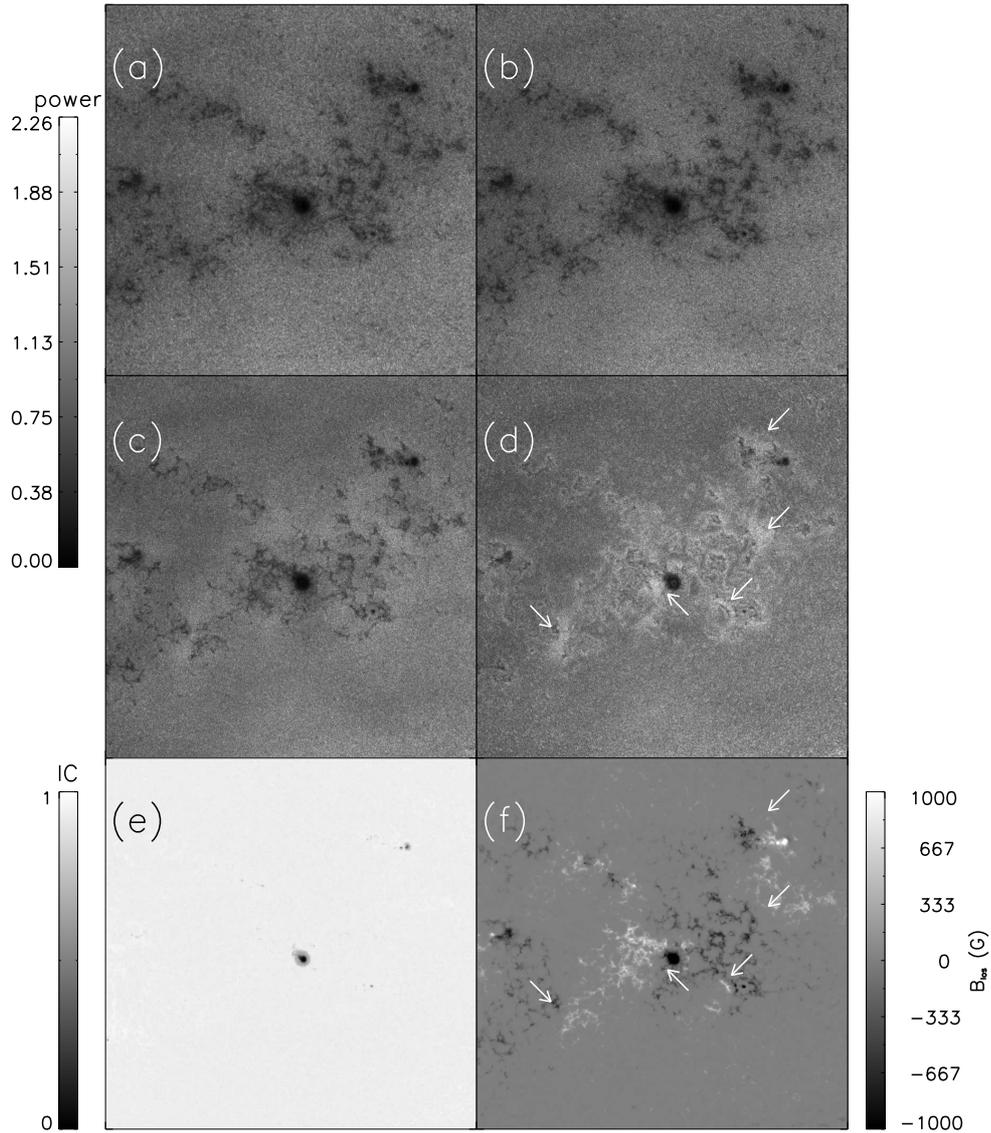}
%power_maps.pro
\caption{24 January 2002 AR 9787 acoustic power maps at (a) 3~mHz, (b) 4~mHz, (c) 5~mHz, (d) 6~mHz,   with the intensity continuum (e) and the  line-of-sight magnetic field (f). The acoustic power maps are normalised to one in the quiet-Sun. The arrows in panel (d) indicate bright halos located between opposite polarity regions as seen in panel (f). The length of each side of the shown observations are 
$\approx 975$~Mm.}
\label{acpow9787}
\end{figure}

The first four panels of Figure~\ref{acpow9787} show acoustic power in 1~mHz bands centered on (a) 3~mHz, (b) 
4~mHz, (c) 5~mHz and (d) 6~mHz for the 24 January 2002. The sunspot in AR 9787 is seen close to the center.  The bottom two panels show intensity continuum (e) and the line-of-sight 
magnetic field (f). In the lower 
frequency bands, (a) and (b), we see suppression of acoustic power coincident 
with the location of magnetic field. In (c) we begin to see an enhancement of 
power surrounding the suppression in plage regions, and in (d) we see 
predominantly enhanced power except in the very strongest of magnetic fields, 
within the sunspot itself.  It is clear that both the suppression (at lower frequencies)
and enhancement (at higher frequencies) are 
spatially correlated with the magnetic field, even in the smallest
plage regions. It is also interesting to note from 
panels (d) and (f) that the strongest 6~mHz enhancements are located between regions 
of opposite polarity (indicated by the arrows), leading us to 
explore the dependence upon magnetic field inclination. 

\begin{figure}
\includegraphics[width=1.0\textwidth]{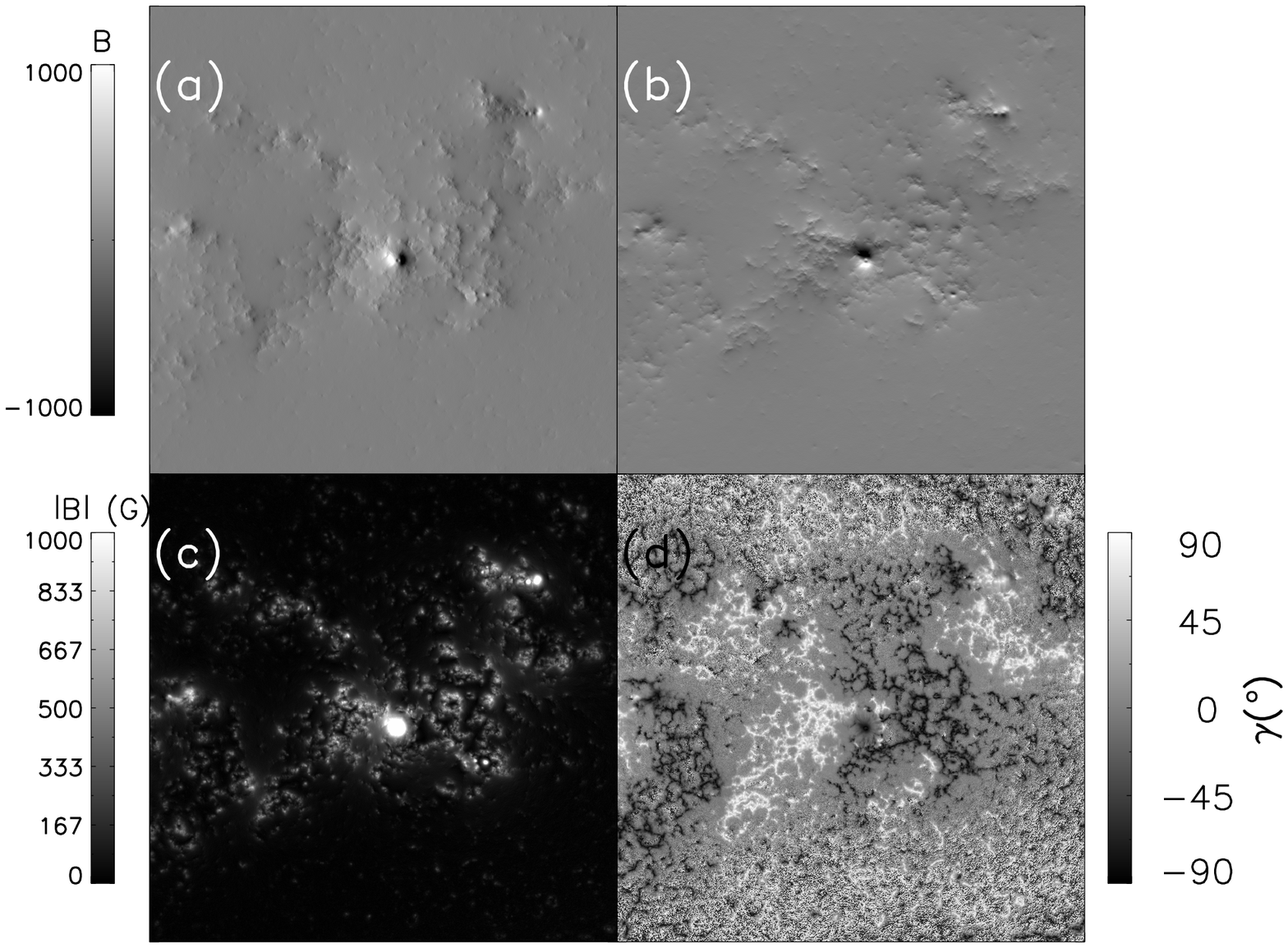}
% mag_maps.pro
\caption{24 January 2002 AR9787 potential field extrapolation (a) $B_x$, (b) $B_y$, (c)  $|B|$,  from 0 to 1000~G and (d) $\gamma$.  $B_z$ is not shown here, but it is the same as the line-of-sight component in Figure~\ref{acpow9787} (f). The length of each side of the shown observations is  $\approx 975$~Mm.}
\label{vecmag}
\end{figure}

\begin{figure}
\begin{center}
%\includegraphics[width=0.9\textwidth]{plot_inclination_power_150B350_020124.ps}
% plot_power_inclination_one.pro
%convert -density 300x300 plot_inclination_power_150B350_020124.ps eps2:pnew.ps
\includegraphics[width=0.9\textwidth]{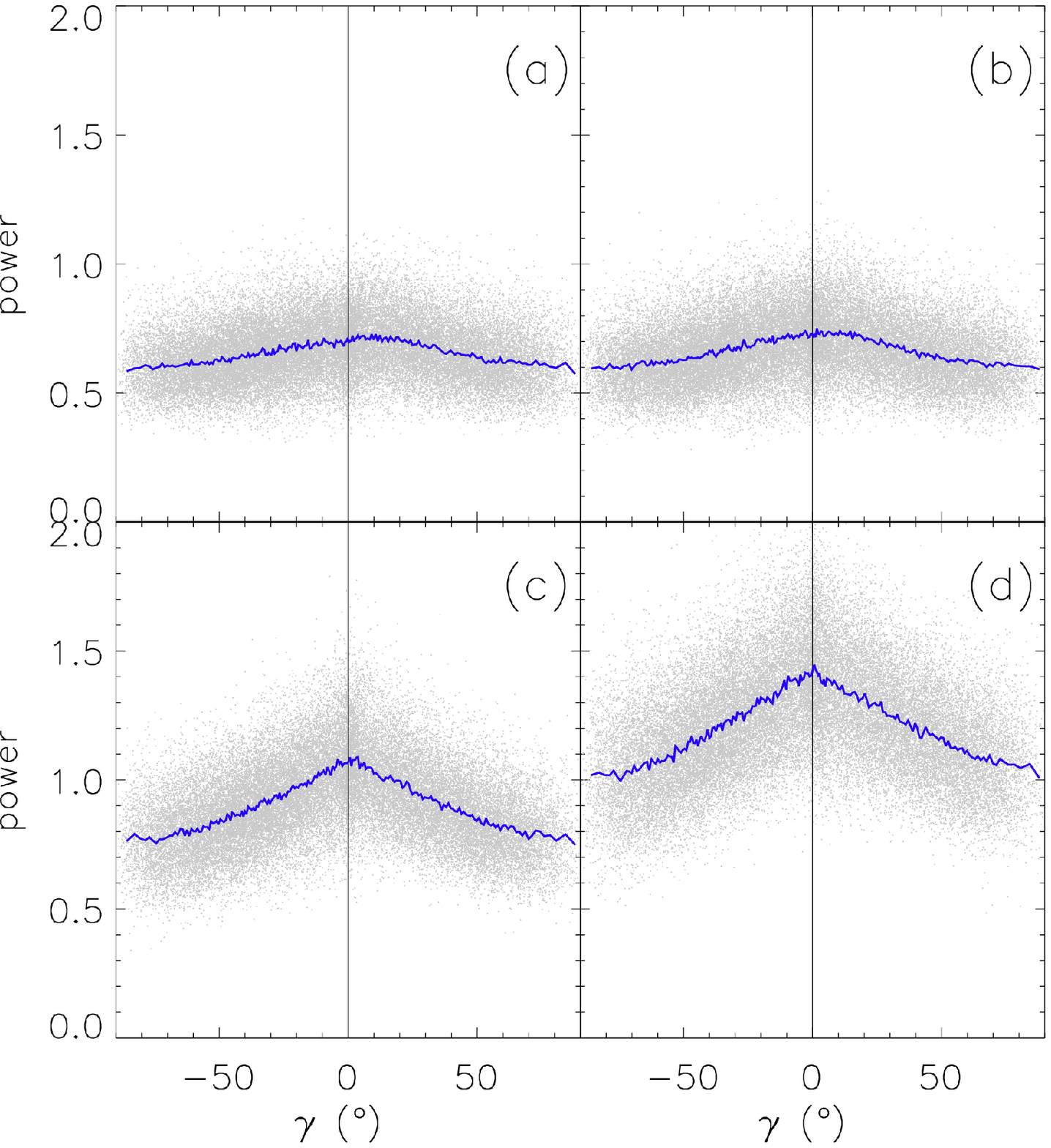}
\caption{Normalised acoustic power against inclination, $\gamma$,  for magnetic field 
strength $150  \le |B| \le 350$~G for (a) 3~mHz, (b) 4~mHz, (c) 5~mHz and (d) 
6~mHz. The blue curve is the binned average of 200 points. The vertical line 
indicates horizontal field.}
\label{scatp}
\end{center}
\end{figure}

Figure~\ref{vecmag} shows the vector magnetic field, (a) $B_x$, (b) $B_y$ and (c) $|B|$ derived from the line-of-sight observations in Figure~\ref{acpow9787} (f). 
From these maps we made scatter plots of the acoustic power 
against the field inclination for each of the frequency bands shown in 
Figure~\ref{acpow9787}. The field 
inclination, $\gamma$, is defined as $\tan( \gamma)=B_z/B_h$ where $B_h=\sqrt{B_x^2 + 
B_y^2}$ is the horizontal field strength and $B_z$ is 
the field parallel to the radial vector. Thus, purely horizontal field is defined when $\gamma = 0^\circ$, $\gamma < 0^\circ$ when $B_z < 0$ and $\gamma > 0^\circ$ when $B_z > 0$.
  Figure~\ref{scatp} shows the results for 
(intermediate) magnetic field strength, $150  \le |B| \le 350$~G, over each of the four frequency bands
centered at 3, 4, 5, and 6~mHz. For all frequency bands, there is a pronounced peak
in power at $\gamma = 0^\circ$. This is true for the lower frequencies,
at which the power in the magnetic regions shows suppressed levels
compared to the quiet-Sun, but is more pronounced at 5~mHz and 6~mHz. 
Acoustic power in these intermediate-strength 
magnetic pixels at 5~mHz can be either suppressed (in more vertical fields) or enhanced
(in more horizontal fields), compared to the quiet-Sun.
At 6~mHz all of these pixels show enhanced power.

\begin{figure}
\includegraphics[width=0.7\textwidth,angle=90]{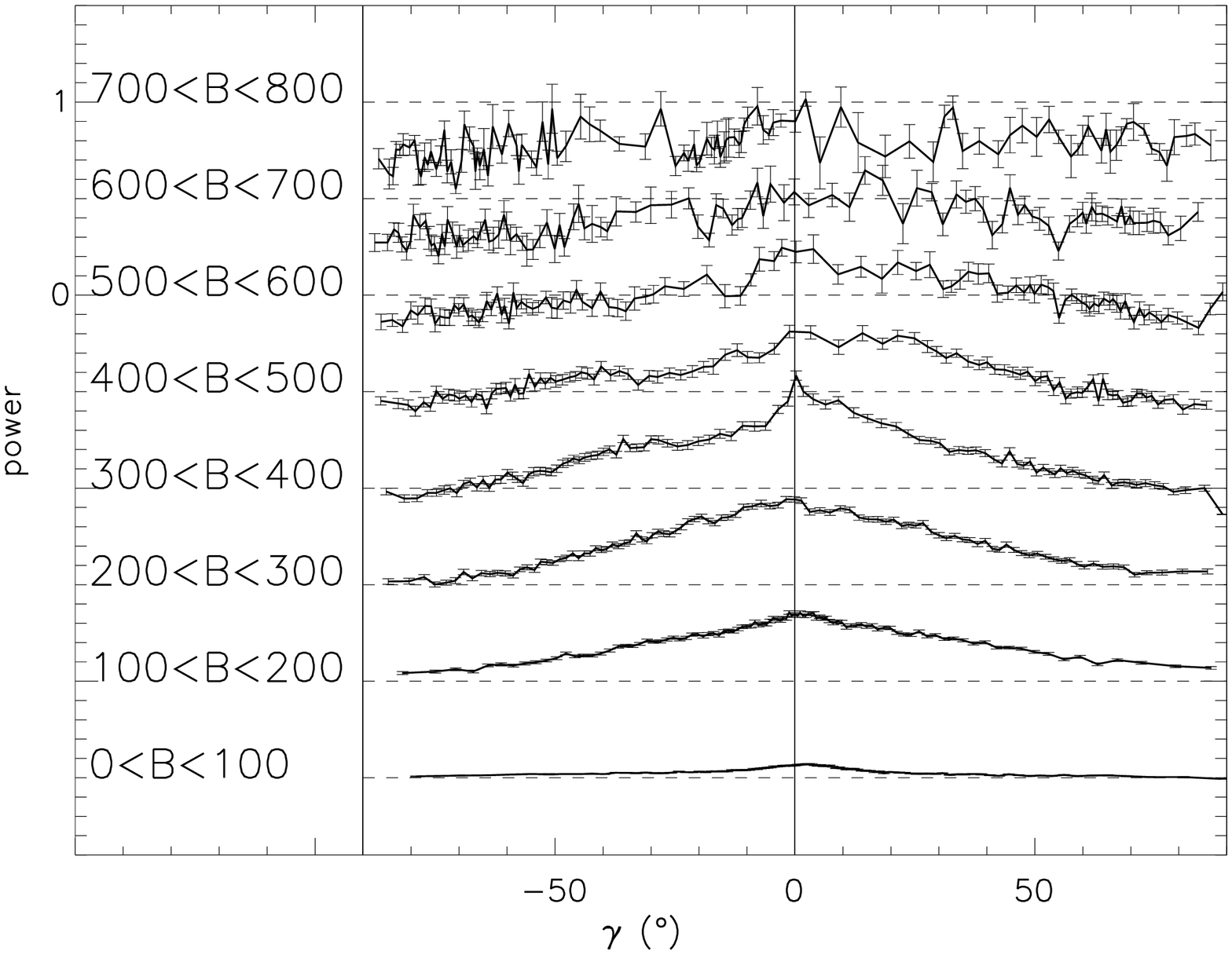}
%plot_power_inclination_one.pro
\caption{Mean 5.5 - 6.5~mHz acoustic power against inclination in ranges of 
magnetic field strength as indicated. The points have been 
averaged in bins of inclination comprising 1\% of the total number of points in each
field-strength range and the standard deviation is over-plotted. Each curve, except the top, has been shifted down for ease of display, where the dashed curves represent a value of 1 for each magnetic field band.}
\label{Bband}
\end{figure}

Figure~\ref{Bband} shows the variation of acoustic power in the 6~mHz bandpass 
with inclination over a wider range of total magnetic field strength. 
For clarity, only the acoustic power, averaged over bins in $\gamma$ comprising 
1\% of the total number of points, 
is shown for each magnetic field strength range.
It is clear that the trend of increased power in horizontal fields is present
for all ranges of field-strength shown, although the results become noisier
at higher strengths due to poorer statistics.
The greatest enhancement (50\% above the quiet-Sun power) 
occurs for field strength between 300 and 400~G, 
and falls off with either greater or lesser strengths.
There is a power suppression at the highest field strengths, occurring 
mostly in the sunspots.

\begin{figure}
\includegraphics[width=1.0\textwidth]{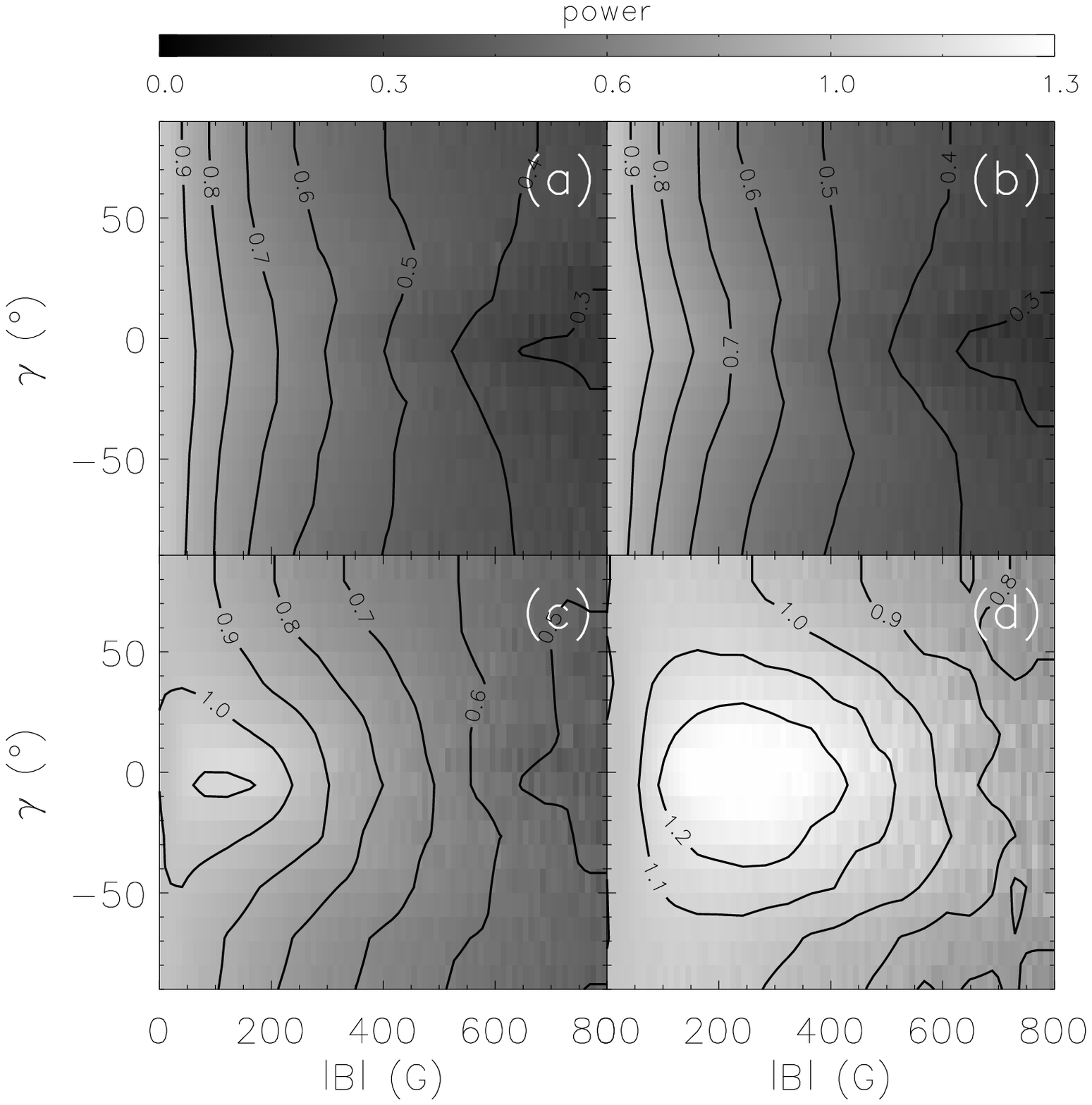}
%squares_ave.pro
%disp_squares_contour.pro
\caption{Mean acoustic power in $10^\circ$ bins of inclination, $\gamma$, and 
in 10~G bins of magnetic field strength in 1~mHz band passes centred at (a) 3~mHz, (b) 4~mHz, (c) 5~mHz and (d) 6~mHz averaged over 7 days (21 - 27 January 2002) where $\gamma=0^\circ$ is horizontal field.}
\label{squares1}
\end{figure}

To visualise the dependence of the normalised power on both the field strength
and inclination, we compute the
average power in 10~G bins of field 
strength and $10^\circ$ bins of inclination. Figure~\ref{squares1} shows the 
2D distribution of the mean normalised power for 1~mHz bands centered on (a) 3~mHz, (b) 4~mHz, (c) 5~mHz and (d) 6~mHz. 
The distribution of enhanced power is symmetric about the horizontal 
field at 5 and 6~mHz, with most of the enhanced power occurring within
a narrow range of inclination and field strength. 

\begin{figure}
\includegraphics[width=1.0\textwidth]{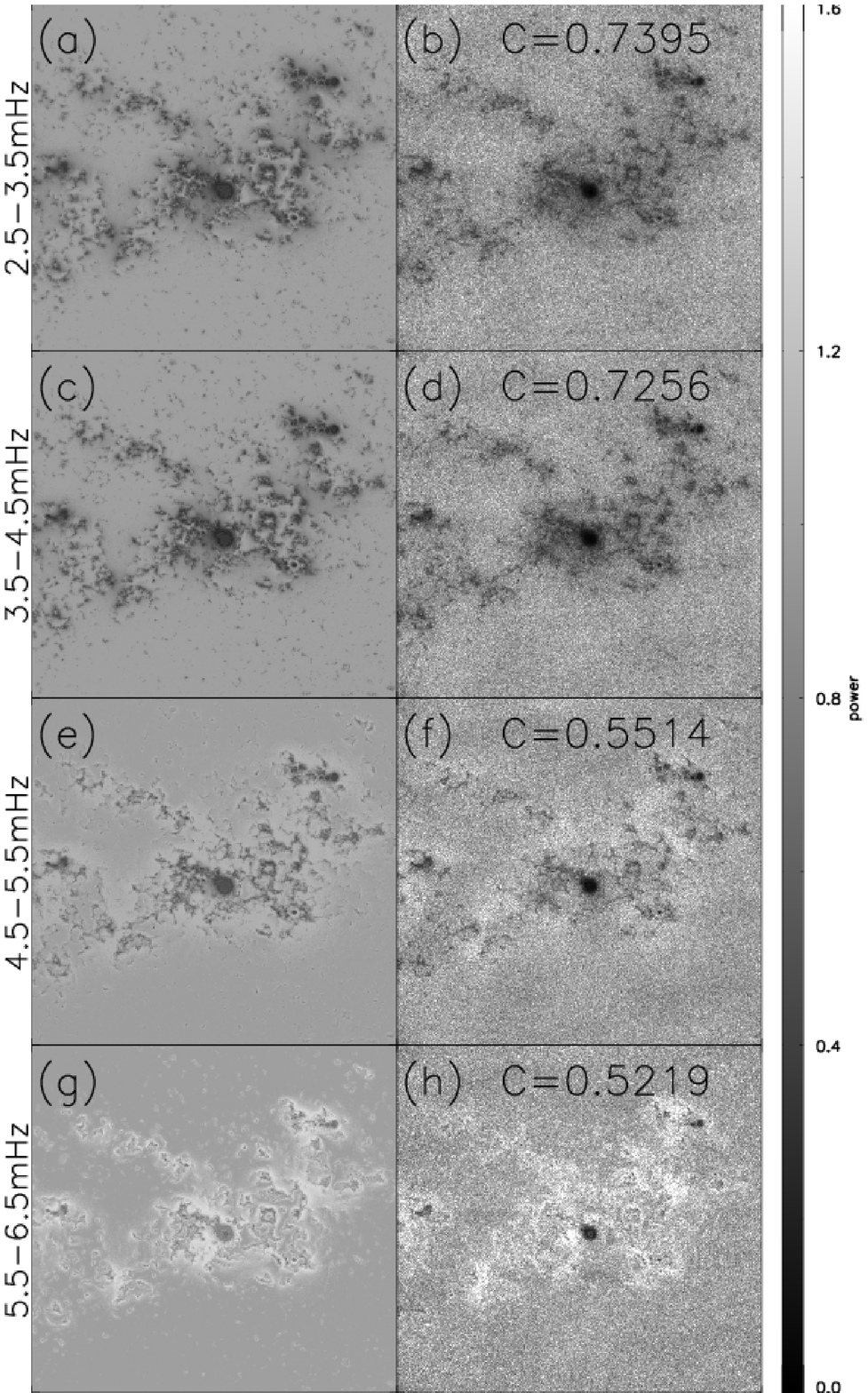}
%%%create_power_map.pro
%% convert -density 100x100 created_power_maps.ps eps2:created_power_maps.eps
\caption{The reconstructed acoustic power maps for 24 January 2002 are shown in the left column and the right column shows the observed acoustic power map. Each frequency band (3~mHz, 4~mHz, 5~mHz and 6~mHz) is shown from top to bottom row. The reconstruction is qualitatively successful with the correlation coefficient, $C$, between the original and reconstructed power maps given on the right-hand-side.  The length of each side of each map is $\approx 975$~Mm.}
\label{cpm}
\end{figure}

We explore the possibility that the functional dependence shown in Figure~\ref{squares1} 
is sufficient to predict the appearance of the acoustic halo using only the extrapolated 
magnetic field components. 
Figure~\ref{cpm} (left column) shows the reconstructed acoustic power maps using
the magnetic field for 24 January 2002 (as shown in Figure~\ref{vecmag}), and the 
 2D distributions illustrated in Figure~\ref{squares1}. The right column of Figure~\ref{cpm}  shows the actual normalised acoustic power map for 24 January 2002. Each row of Figure~\ref{cpm} represents a different frequency band.  The reconstruction is qualitatively successful with the correlation coefficient, $C$, between the maps given on the right-hand-side. The correlation is best for low frequencies. Recall that these maps are created only for the magnetic field map of \textit{one} day (24 January 2002) but using Figure~\ref{squares1} which is an \textit{average} of power over all seven of the observed days. Even so, the correlation coefficient for using power from the same day is very similar ($C$ is 0.74, 0.73, 0.55, 0.53 for 3, 4, 5 and 6~mHz bands respectively).
It therefore appears that the acoustic halo is largely a property of the 
local surface magnetic field.

\begin{figure}
\includegraphics[width=0.47\textwidth]{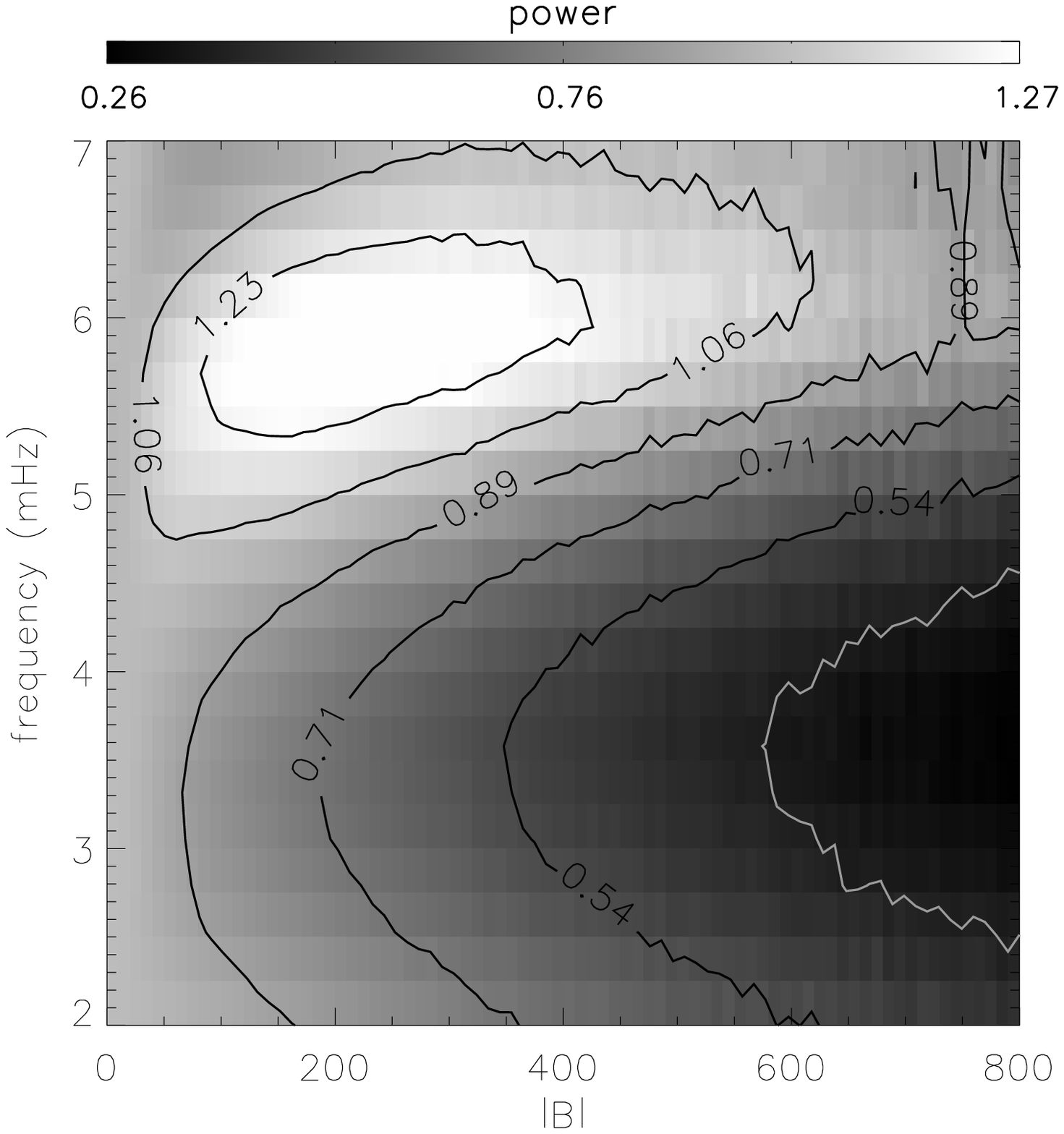}
\includegraphics[width=0.47\textwidth]{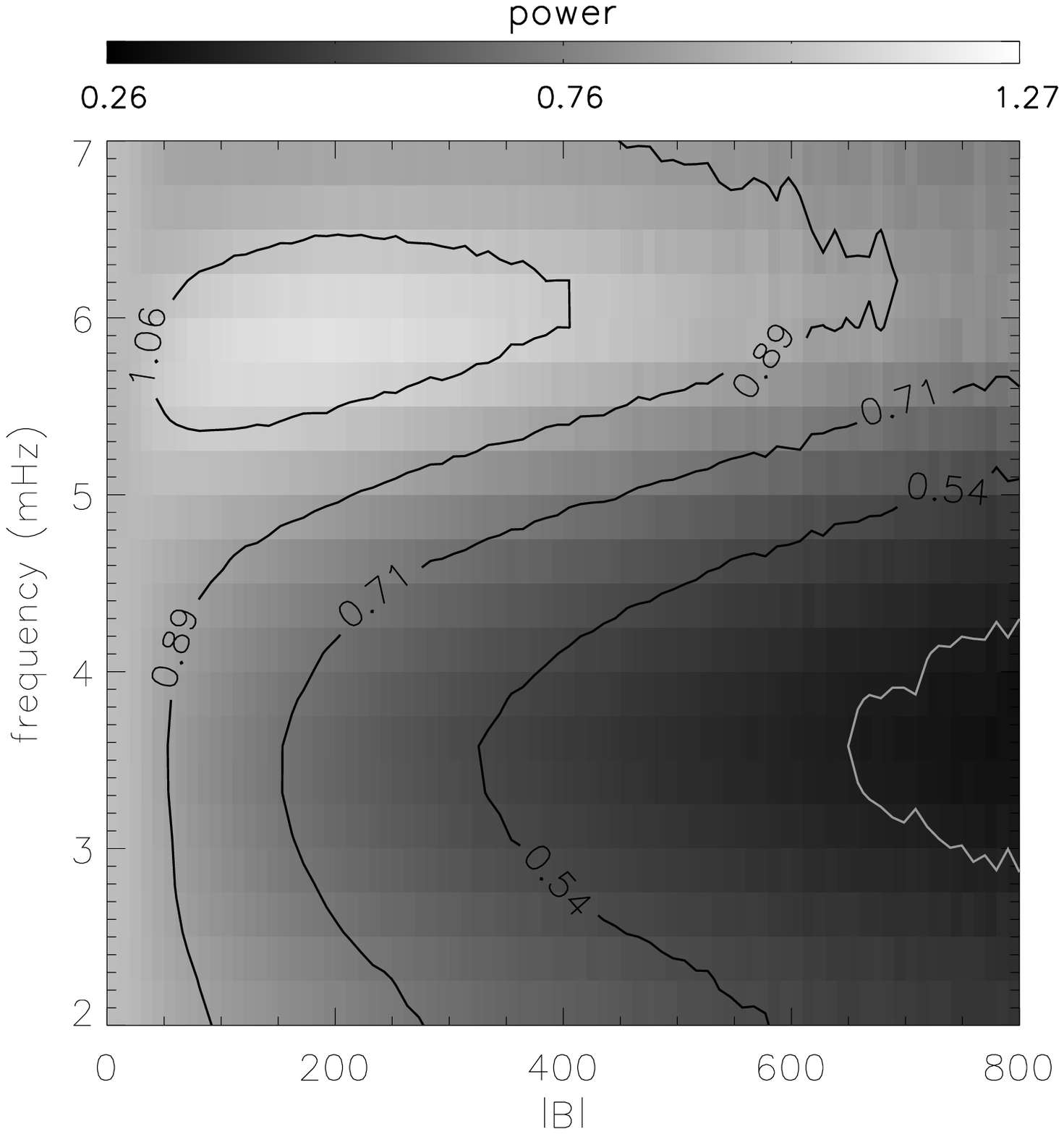}
%PRO/squares_0G30.pro
%PRO/squares_30G90.pro
\caption{Mean acoustic power in 0.1~mHz bins of frequency and in 10~G bins of 
magnetic field strength averaged over 7 days (21 - 27 January 2002) for 
$0^\circ \le | \gamma | < 30^\circ$ (left) and $30^\circ \le | \gamma | < 
90^\circ$ (right). The lowest power contour is lighter in colour for visibility. }
\label{squares_freq}
\end{figure}

There is some indication from Figure~\ref{squares1} that the peak in the
mean normalised power at 6~mHz occurs at a higher field strength than the
peak in the distribution at 5~mHz. To explore this variation further, we
construct a new set of normalised acoustic power maps with narrower
frequency bandpasses.
From the symmetric properties of the acoustic power around  
$\gamma=0^\circ$ we 
can comfortably work with $| \gamma |$. Figure~\ref{squares_freq} shows the 
mean acoustic power in 10~G bins of magnetic field and 0.1~mHz bins of 
frequency for $0^\circ \le | \gamma | < 30^\circ$ (left) and $30^\circ \le | 
\gamma | < 90^\circ$ (right).  A clear trend of increasing peak frequency
with $| B |$ is evident in both ranges of inclination.
The well known suppression of power at 3--4~mHz, which is stronger at  
greater field strength, also shows up prominently in Figure~\ref{squares_freq}.

\section{Wavenumber dependence}\label{two}

We now turn to examining the nature of the enhanced power oscillations in terms 
of changes in the frequency and wavenumber domain. 
To isolate the power spectra of the halo regions we apply spatial masks 
to the observations before performing a 3D Fourier transform. Since the power spectra are
quite sensitive to the window function, or mask, it is necessary to compare the
power spectra of observations of the halos with those of quiet-Sun using identical
masks.  Because of the large filling factor of the active regions in the
2002 January data this comparison nominally requires an additional 
7 days of observations of predominantly quiet-Sun. This 
entails using additional MDI observations from a substantially quiet epoch.
For the quiet-Sun comparisons used here, we chose observations (hereafter
referred to as the ``2003 observations") from 24 to 27 July 2003 and 18 to 20
October 2003.  We refer to the 7 days of data for the active period  
from  21 to 27 January 2002 as the ``2002 observations''. 
We track and Postel-project the 2003 observations in an identical manner to the 
2002 observations and then pair the 2002 and 2003 observations in days according to 
Table~\ref{table1}. 

Unfortunately, focus changes in the MDI instrument were made between the 2002 and 2003 
observations. In addition, the time span between the two datasets is substantial enough
that possible solar cycle changes in mode properties could confuse our power spectra
comparisons.
Thus it is necessary to look for, and rule out, possible changes in the
power spectra of the quiet-Sun over the same two epochs which might be due to either
instrumental or solar causes.
We therefore create two sets of masks; the first of which (called the ``halo masks'')
is intended to isolate the
halo regions in the 2002 observations, and the second (``quiet-Sun masks'')  is intended 
to isolate the quiet regions in the 2002 observations. 
An additional complication is that the 2003 observations
also include magnetic regions which must be eliminated from the
comparison. To do this, the halo masks for each pair of days 
include only pixels with  
$100 < |B| < 350$~G and $|\gamma | \le 20^\circ$ in the 2002 
observations and with $|B| \le 30$~G in the 2003 observations.  
The set of quiet-Sun masks is constructed for each pair of days
from pixels with $|B| \le 30$~G in both the 2002 and 2003 observations.
This excludes any magnetic pixels present in the 2003 
observations, and ensures that the masks used for both sets of days are identical.

\begin{figure}
\begin{center}
\includegraphics[width=0.8\textwidth]{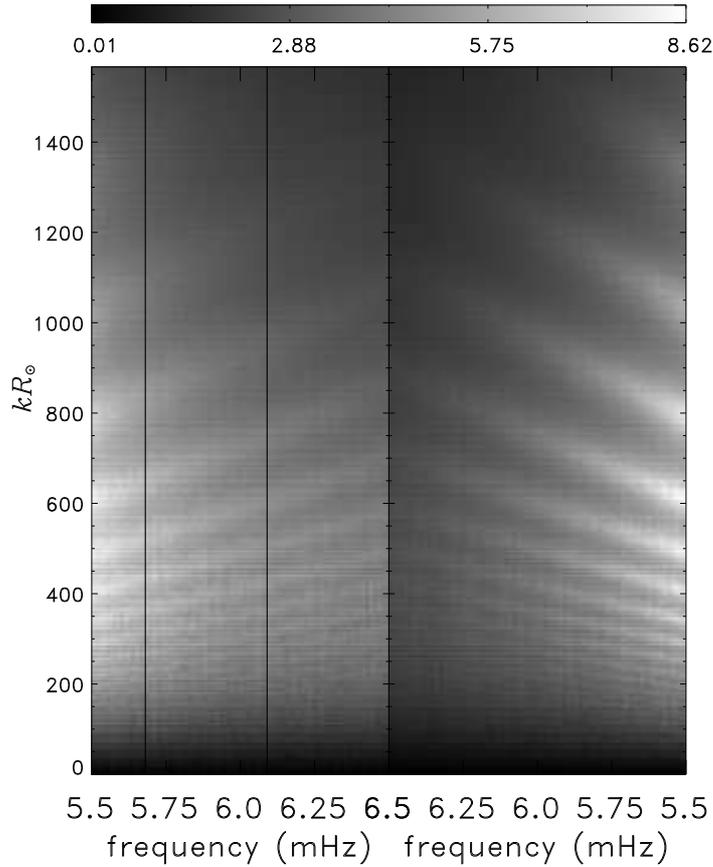}
%/data/seismo/schunker/for_doug/make_compare_ps.pro
\caption{The azimuthally summed power spectra (in arbitrary units) for the halo masks applied to the
2002 (active region) observations and averaged over all days (right panel) and for the same masks
applied to the 2003 (quiet-Sun) observations averaged over all days 
(left panel). The vertical lines correspond to the cuts plotted in Figure~\ref{pcuts}.}
\label{spectra}
\end{center}
\end{figure}

Using these masks we compute the azimuthally summed power spectrum for each pair of
days. The 7-day average of the 2002 and 2003 spectra are shown in Figure~\ref{spectra}.
It is apparent that the power in the 2002 spectra of the halo regions is 
greater, and the ridges are more defined, than the spectra in 2003.
Moreover, there appears to be a shift towards higher $k$ 
in the mode ridges in the halo power spectra which is particularly visible at
higher temporal frequency.

\begin{figure}
\hspace{1cm}
\includegraphics[width=0.9\textwidth]{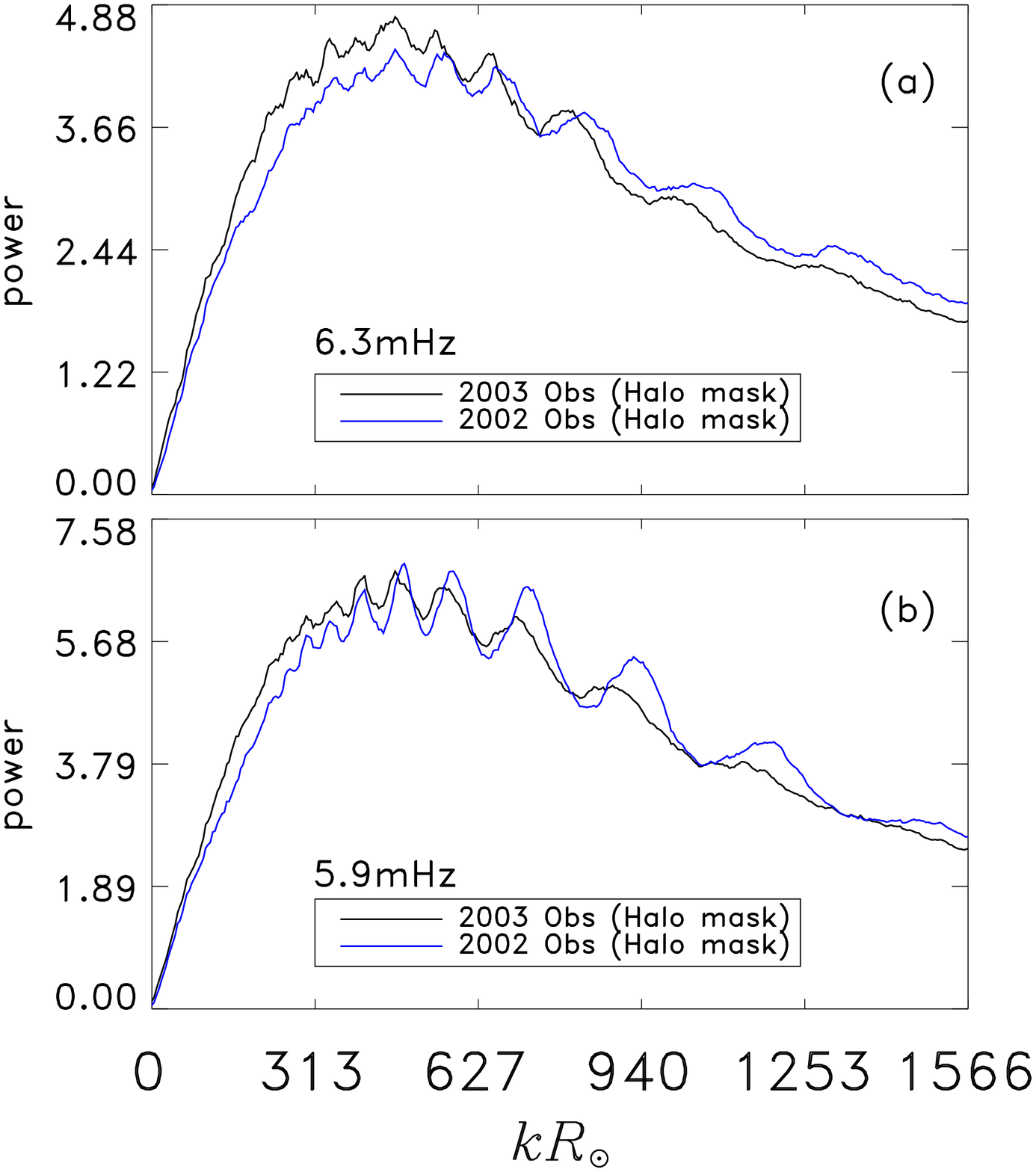}
\vspace{0.5cm}
%/data/seismo/schunker/for_doug/make_compare_ps.pro
\caption{Cuts through the power spectra (in arbitrary units) shown in Figure~\ref{spectra} at 
constant frequency. The top panel (a) shows a cut through the power spectrum at 6.3~mHz for 
the halo-masks applied to the 2002 (active region) observations (blue) and a cut through the 
the 2003 (quiet-Sun) power spectrum (black). The same is 
shown in the bottom panel (b) except at a frequency of 5.9~mHz. At 5.9~mHz, the 2003  
(quiet-Sun) power spectrum has been boosted by 1.21 to allow an easier comparison with the 2002 
spectrum. Some smoothing of the curves was performed for the purpose of the plot.}
\label{pcuts}
\end{figure}

To show this more clearly, we plot cuts through the spectra in Figure~\ref{pcuts} at 6.3~mHz 
(panel a) and at 5.9~mHz (panel b), as indicated by the vertical lines in the 
power spectra (Figure~\ref{spectra}). 
To facilitate the comparison at 5.9~mHz the quiet-Sun cut has been 
multiplied by a factor of 1.21.
From these plots it is clear that there is a shift of the ridges in 
the 2002 halo-masked power spectrum to higher $k$. 

\begin{table}
\caption{The dates of observations for pairing the 2002 observations (January) 
 and the 2003 observations (July and October).}
\label{table1}
\begin{tabular}{| c | c | c | c | c | c | c | c | c | c |}
\hline  	
  \textbf{2002 obs}    & Jan & 21 & 22 & 23 & 24 &   & 25 & 26 & 27 \\
  \hline
  \textbf{2003 obs}  & Jul & 24 & 25 & 26 & 27 & Oct & 18 & 19 & 20 \\
\hline  
\end{tabular}
\end{table}

\begin{figure}
\hspace{1cm}
\includegraphics[width=0.9\textwidth]{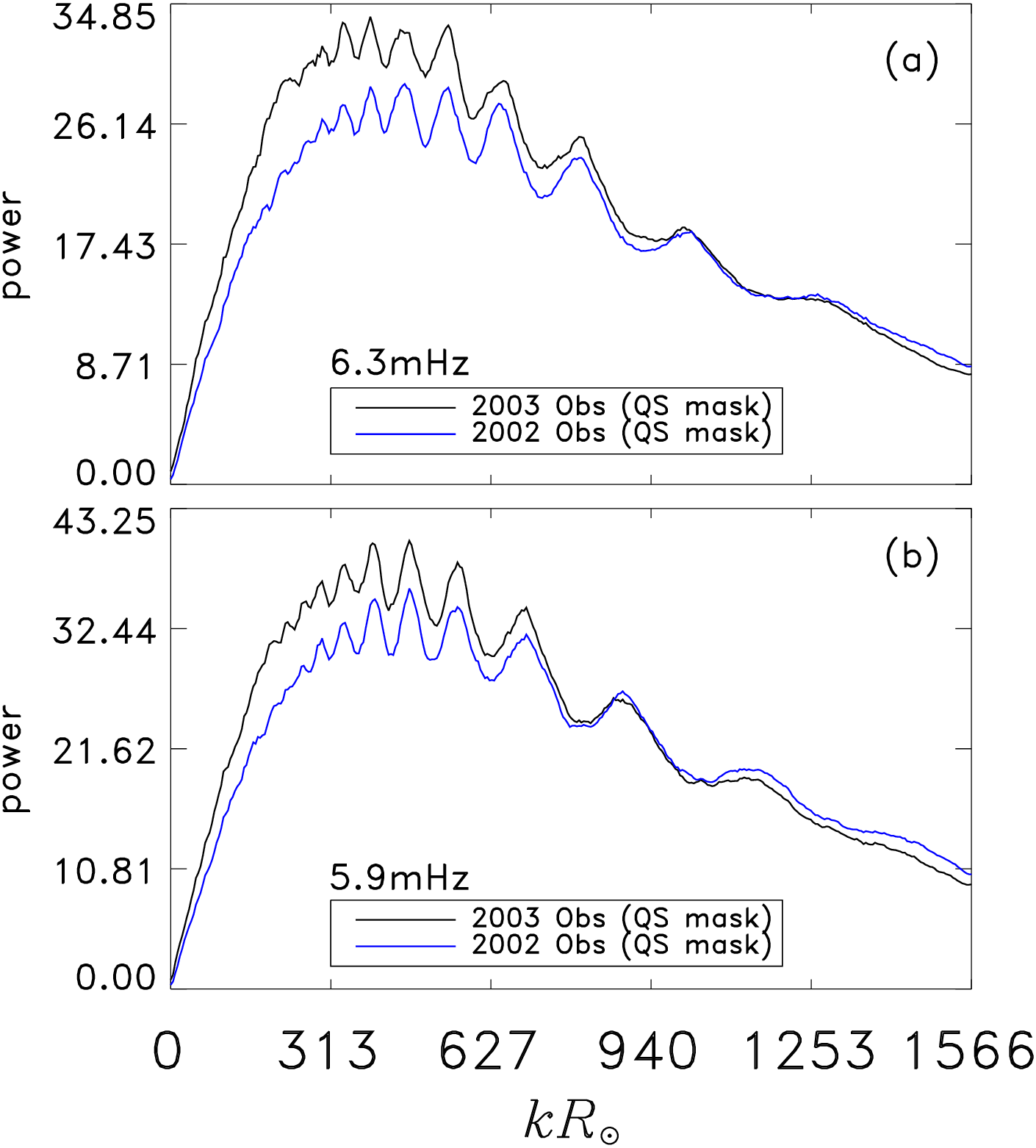}
\vspace{0.5cm}
%/data/seismo/schunker/for_doug/make_compare_qs.pro
\caption{The same as Figure~\ref{pcuts}, except the cuts are taken through the 
power spectra after applying the quiet-Sun mask set.
The top panel (a) shows smoothed power 
cuts at 6.3~mHz for quiet-Sun regions in the 2002 observations (blue) and quiet-Sun 
regions from the 2003 observations (black). The same is shown in the bottom panel 
(b) except at a frequency of 5.9~mHz. }
\label{qscuts}
\end{figure}

Similar shifts of the ridges
between the 2002 and 2003 quiet-Sun regions, if present, 
would indicate some artifact
or other effect which is not unique to the halos.
Figure~\ref{qscuts} shows cuts through the power spectra obtained
with the  quiet-Sun masks. It is apparent that
some differences in the magnitude and distribution of power, but not
frequency shifts, between the two
epochs are observed. The former is not unexpected 
given the known focus changes between the two epochs. 
However, it is clear that we do not see a shift in the position of the 
ridge peaks at constant 
frequency as is seen in Figure~\ref{pcuts}. This strengthens the likelihood 
of real differences in the modal properties between halo and quiet regions. 

\section{Discussion}\label{discussion}
We find properties of the acoustic halo that confirm prior analyses, but also
discover several heretofore unknown properties.
The halos are locations of acoustic power increase up to 140\% 
of the quiet-Sun value at higher frequencies and are strongest at intermediate magnetic field
strength (150 to 350~G).  These results are in general 
agreement with previous studies \cite{JH02,Hindman:1998p398}. We have 
discovered a clear association with near horizontal ($\pm 30^\circ$) magnetic field. The dependence of acoustic amplitude on field inclination occurs
over frequencies between 2.5 and 6.5~mHz, and for field strengths between 100 and
800 gauss, but is particularly strong between 5 and 6.5~mHz and at the aforementioned
intermediate field strengths.  In addition, we show that the frequency of the peak enhancement depends on the field strength, with the peak frequency increasing slightly with field strength, and that in these regions of enhanced power the ridges of the  acoustic power spectrum are shifted towards higher $k$.  
Qualitatively there appears to be a larger shift for higher $k$ indicating that this is a shallow surface effect. For constant $k$ the phase speed of the ridge is reduced, indicating an increase in travel times, that could be caused by either a slower wave speed  or a longer path distance.

Any theory attempting to explain the acoustic halo or fully explain the interacation of waves with the magnetic field needs to 
incorporate these properties. A quantitative study of changes in the power spectrum 
requires a much more in depth analysis than our rudimentary study discussed here
since, amongst other things, we have made no attempt to remove or account for 
distortions due to 
foreshortening and projection effects from the power spectra. 
On the other hand, we are confident through the multiple comparisons 
performed here with different masks and solar regions that the ridge shifts 
are qualitatively real and are clearly associated with the halo regions. 

Various authors have pointed to a few different mechanisms to explain the halo.
Some of the original discovery papers suggested enhanced acoustic emission
\cite{BDLJHP92, BBLT92}.  
More recently, \inlinecite{JKWM08} have studied the effects of vertical magnetic fields on solar convection. The convection occurs on smaller scales and higher frequencies than in quiet-Sun regions, and shifts the oscillation power to higher frequencies. It is not as yet understood whether these properties are dependent upon the inclination of the field.
However, the lack of halos in continuum observations \cite{Hindman:1998p398} and the relative lack of association with emission sites, as identified with seismic holography analyses sensitive to propagating waves \cite{Donea:2000p120} casts doubt on this interpretation.

\inlinecite{Hanasoge:2008p816} claims that halos may 
be more likely related to scattering effects. Roughly in agreement with the findings 
in this paper, he also sees the halo as being confined to magnetic field 
inclinations $|\gamma| < 40^\circ$ and an increase of energy in the higher wavenumber regime. \inlinecite{KC2009}  propose that 
high-frequency fast mode waves refracted in the higher atmosphere due to a 
rapid increase in the Alfv\'en speed are responsible for the seismic halos. 
They get a 40-50\% increase in power of high frequency vertical velocity in 
regions of intermediate field strength, although, like 
\inlinecite{Hanasoge:2008p816}, they find that the horizontal velocity 
component of the halo is stronger. However, it is difficult to completely reconcile the observations with these 
simulations, since the latter show only a weak halo being generated in 
the vertical velocity component (up to a maximum of 5\%) and 
the strongest enhancement (up to 35\%) being generated  in the horizontal 
velocity component.  If the halo were strongest in horizontal velocities 
there would be a line-of-sight dependence with the halo power being 
stronger at larger heliocentric distance. This has not as yet been observed. 

\inlinecite{Kuridze:2008p578}  suggests this 
enhanced power is caused by high frequency waves trapped  beneath the
bipolar magnetic canopy. The low frequency waves are reflected below 
the photosphere and so are not trapped. In regions of open field, the high 
frequency waves are permitted to propagate upwards and are lost to the 
atmosphere.  These field free cavities have granular dimensions ($\approx 
0.5$~Mm), but with SOHO/MDI we must be observing \textit{many} of these 
cavities within the resolution which is three times this size, and these 
cavities must be widely distributed to cover the area of the halo regions. 
A somewhat similar idea is put forth by  \inlinecite{MHS05}. They suggest that the lack of haloes in the TRACE observations \cite{M05} caused by reflections (and mode conversion) due to an overlying inclined magnetic field. 

\inlinecite{Simoniello:2009} have studied Sun-as-a-star oscillations and 
observe enhanced power in high frequency (5.7-6.3~mHz) bands, up to 18\%, at 
solar maximum. We believe that this is due to a large surface of the Sun being 
covered by plage and the associated enhanced power, as presented in this paper, 
associated with the increased magnetic activity. 

Understanding the properties of surface acoustic power is important in modeling 
helioseismic signatures due to subsurface perturbations.
It has also been suggested that acoustic power be used as a predictive tool to forecast the emergence of active regions 
\cite{HKZM09}. 
The results shown here, namely that the surface acoustic power is a strong function of the local
magnetic field properties, may be useful in the removal of surface contributions
from acoustic power maps and as a part of a correction to local helioseismic methods
(e.g. \inlinecite{LB05},\inlinecite{CR07}).

At this point, we cannot confidently suggest a mechanism explaining all the 
properties of the acoustic halo. 
However, a more detailed analyses of the characteristics of the acoustic halo will likely help
to pin-point the mechanism behind the acoustic halos. 
The recently launched Solar Dynamics Observatory hosts the Helioseismic and Magnetic Imager instrument 
which has a resolution four times better than SOHO/MDI and, along side improvements in numerical modeling,
may be able to help resolve the nature of acoustic halos. 

\acknowledgements
We would like to thank Aaron Birch and Ashley Crouch for helpful comments on the manuscript. HS is supported by   the European Helio- and Asteroseismology Network (HELAS), a major international collaboration funded by the European Commission's Sixth Framework Programme.
DCB is supported by NASA contracts NNG07EI51C and NNH09CE41C and
NSF grant AST-0406225.

     % format of references provided by the journal (.bst)
\bibliographystyle{spr-mp-sola}

     % name your Bibtex file containing your references (.bib)
\bibliography{papers_bib}
%\bibliography{stuff}
%{sola_ac_pow_bib}  

     % Checking: look if the file containing the ``\bibitem'' exits
     %           so check if the .bbl file exist (bibTeX compilation)
\IfFileExists{\jobname.bbl}{} {\typeout{}
\typeout{****************************************************}
\typeout{****************************************************}
\typeout{** Please run "bibtex \jobname" to obtain} \typeout{**
the bibliography and then re-run LaTeX} \typeout{** twice to fix
the references !}
\typeout{****************************************************}
\typeout{****************************************************}
\typeout{}}

\clearpage

\end{article} 
\end{document}